\documentclass[prb,aps,twocolumn,groupaddress]{revtex4-1}
\usepackage{latexsym}
\usepackage{graphicx}
\usepackage{verbatim}
\usepackage{multirow}
\usepackage [english]{babel}
\usepackage [autostyle, english = american]{csquotes}
\MakeOuterQuote{"}
\usepackage{amsmath}
\newcommand{\beq}{\begin{eqnarray}}
\newcommand{\eeq}{\end{eqnarray}}
\usepackage{mathrsfs}
\usepackage{float}
\usepackage{caption}
\usepackage{bm}
\usepackage[usenames, dvipsnames]{color}
\usepackage{tabularx}

\usepackage{mathtools}
\usepackage{slashed}
\usepackage{physics}	
\usepackage{graphicx}   
\usepackage{epstopdf}
\usepackage{subfigure}  
\usepackage{hyperref}   
\usepackage{bbold}
\usepackage{wasysym}
\usepackage{amsmath}

\begin{document}

\title{Families of Gapped Interfaces Between Fractional Quantum Hall States}
\author{Julian May-Mann}
\affiliation{Institute of Condensed Matter Theory, Department of Physics, University of Illinois at Urbana-Champaign, Urbana, Illinois, USA}
\author{Taylor L. Hughes}
\affiliation{Institute of Condensed Matter Theory, Department of Physics, University of Illinois at Urbana-Champaign, Urbana, Illinois, USA}

\begin{abstract}
Some interfaces between two different topologically ordered systems can be gapped. In earlier work it has been shown that such gapped interfaces can themselves be effective one dimensional topological systems that possess localized topological modes in open boundary geometries. Here we focus on how this occurs in the context of an interface between two, single-component Laughlin states of opposite chirality, and with filling fractions $\nu_1=1/p$ and $\nu_2=1/pn^2$. While one type of interface in such systems has been previously studied, we show that allowing for edge reconstruction effects opens up a wide variety of possible gapped interfaces depending on the number of divisors of $n.$  We apply a complementary description of the $\nu_2=1/pn^2$ system in terms of Laughlin states coupled to a discrete gauge $\mathbb{Z}_n$ field. This enables us to identify possible interfaces to the $\nu_1$ system based on complete or partial confinement of this gauge field. We determine the  tunneling properties, ground state degeneracy, and the nature of the non-Abelian zero modes of each interface in order to physically distinguish them.
\end{abstract}
\maketitle

\section{Introduction}
Topological phases of matter in two dimensions are often characterized by a gapped bulk interior in addition to gapless edge modes. Phases with topological order can exhibit a number of interesting properties such as quasiparticles with fractional statistics, topological ground state degeneracy on manifolds with non-zero genus, and chiral edge modes\cite{Wen1}. However, having gapless edge or interface modes is not a necessary requirement for topological order, and some systems can support gapped edges with the vacuum or, in general, gapped interfaces to a system with a different topological order\cite{HaldaneNV,kapustin2011,kitaev2012,Wen10,Levin2,kapustin2014}. Indeed, some recent work has explored heterogeneous interfaces between two topologically ordered phases and illustrated that the interface can support non-Abelian bound states, and unusual entanglement properties\cite{cheng2012,lindner2012,clarke2013, vaezi2013,barkeshli2013,mong2014,khan2014,cano2014,santos2016,khan2017,fliss2017,santos2018}.

In this article we focus on interfaces in systems with intrinsic Abelian topological order described within the K-matrix formalism\cite{Wen8}. We revisit the work carried out in Ref. \onlinecite{santos2016} that described heterogeneous interfaces between two one-component topological phases, and uncover new types of gapped, topological interfaces that were not discussed earlier. We consider each interface type in turn, focusing on tunneling properties, ground state degeneracy, and the existence of non-Abelian bound states. These properties are consistent with the perspective that the gapped interfaces we are studying can be interpreted as 1d topological systems similar to parafermion wires\cite{kitaev2001,bondesan2013,motruk2013,fendley2014,jermyn2014,zhuang2015,alexandradinata2016} as was previously discussed in Refs. \onlinecite{santos2016,santos2018}. 


Within the $K$-matrix formalism it is straightforward to determine the families of gapped edges of a topological phase  using one of two equivalent methods: the null vector criteria\cite{HaldaneNV} or the Lagrangian subgroup criteria\cite{Levin2,Qi3}. These methods can also be applied to gapped interfaces between two topologically ordered materials, since we can imagine folding the heterogeneous interface to form a single edge with multiple components. Here we will consider a particularly simple interface between two, one-component Laughlin states. The ideas presented here could be generalized to other interfaces, although the details may be more complicated. The one-component Laughlin states have a single chiral/anti-chiral edge channel and we will focus on interfaces between $1/p$ and $1/pn^2$ states; the fractions are chosen precisely so the interface can be gapped\cite{Levin2}. It has been shown that these two states are related by gauging a $\mathbb{Z}_n$ symmetry of the $1/p$ state, or, conversely, by confinement of the $\mathbb{Z}_n$ charged particles of the $1/pn^2$ state \cite{Ashvin1,khan2017}, and we will exploit this relationship below. Despite the simplicity of the topological orders at the interface, we will show that there are actually a variety of possible interfaces and phenomena which can occur here. 

Our article is organized as follows. We first describe three inequivalent, gapped interfaces between the two topological orders in terms of the (bulk) K-matrix and (edge) Luttinger liquid formalisms. We then heuristically discuss the types of anyon transmission and reflection processes that occur at the various interfaces. From there we move on to discuss the ground state degeneracy of the folded interfaces in disk and cylinder geometries. Finally, we classify the types of non-Abelian modes that appear when a gapped interface intersects a boundary to the vacuum.  

\section{Single Component Interface}
We will begin with a short introduction to the formalism that will describe the two, one-component Laughlin states, and a review of the gapped interface discussed in Ref. \onlinecite{santos2016}. Let us start with the K-matrix theory for the Laughlin $1/p$ ($1/pn^2$) state. For single-component states, the K-matrix of each state is given by the single integer $p$ ($pn^2$). This integer determines the statistics of the $p$ ($pn^2$) types of anyon quasiparticles, as well as the ground state degeneracy $|p|$ ($|pn^2|$) on a torus. From this bulk K-matrix we can construct a Lagrangian for the edge modes, say along the $y$-direction, of an Abelian topological order described by $K$: 
\beq
\mathcal{L} = \frac{1}{4\pi} \partial_t \bm{\phi}^T K \partial_y \bm{\phi} - \frac{1}{4\pi} \partial_y \bm{\phi}^T V \partial_y \bm{\phi}\label{eq:KLL}
\eeq
where $V$ is the velocity matrix, and and $\phi$ is a vector of bosonic fields. The anyons are given by $e^{i \bm{r}\cdot \bm{\phi}}$, where $\bm{r}$ is an integer valued vector. The anyons are indistinguishable upon attachment of local particles given by $e^{i \bm{R}^T K \bm{\phi}}$, where $\bm{R}$ is an integer valued vector. The anyons thereby form a discrete lattice\cite{read1990,wen1992,frohlich1994,frohlich1994,cano2014}.
The $K$ matrix description is invariant under transformations $K \rightarrow W^TKW$, $\phi \rightarrow W^{-1}\phi$, where $W$ is integer valued, has unit determinant, and is the same dimension as $K$, i.e., $W$ is a basis change that leaves the lattice of quasiparticles unchanged.

For the $1/p$ and $1/pn^2$ states respectively we will use the $K$ matrices\cite{Wen1}
\beq
K_p = p\\
K_{pn^2} = pn^2.
\eeq
For the interface, where we have an anti-chiral $1/p$ state abutting a chiral $1/pn^2$ state, we will use the representation 
\beq 
K_t = K_{-p} \oplus K_{pn^2} =  \begin{bmatrix}
    -p  & 0   \\
    0  & pn^2
\end{bmatrix},
\eeq 
with bosonic fields $\bm{\phi}_t = (\phi_{t,1},\phi_{t,2})$. The field $\phi_{t,1}$ corresponds to the edge excitations of the anti-chiral $1/p$ state, and $\phi_{t,2}$ corresponds to the edge excitations of the chiral $1/pn^2$ state. The overall chirality of the interface is determined by the signature of $K,$ which is vanishing in our case, so we have the possibility of gapping the interface. To find terms that can be added to Eq. \ref{eq:KLL} to introduce a gap on the edge, we will use the null vector criteria. For a $2N\times 2N$ $K$ matrix, we must find $N$ $2N$-dimensional vectors $\Lambda_i$ such that $\Lambda_i^TK\Lambda_j = 0$, for all $i$ and $j$ \cite{HaldaneNV}. To prevent any spurious degeneracy occurring at the interface due to spontaneous symmetry breaking, we also require that the null vectors are primitive. This means that the greatest common denominator of all $N\times N$ minors of the $N\times 2N$ matrix $M = [\Lambda_1 ... \Lambda_N]$, is 1 \cite{Levin1}.
From these null vectors we can construct backscattering interactions that entirely gap out the edge: 
\beq
H_I = \sum_i \lambda \cos(\Lambda^T_iK\phi).
\label{eq:gapGe}
\eeq
The null vector criteria guarantees that these backscattering terms commute with each other, and hence can simultaneously gap all the edge channels. 

For our particular choice of interface given by $K_t$, the null vector criteria and primitivity condition for this system are satisfied by $\Lambda_t = (n,1)$. We could have chosen $\Lambda_t = (\pm n,\pm 1)$, but for our purposes this difference is not consequential.
This null vector gives the following backscattering term 
\beq
\nonumber H_I &=& \lambda \cos(\Lambda_t K_t \bm{\phi}_t)\\
&=&\lambda \cos(pn \phi_{t,1} - pn^2 \phi_{t,2}).
\label{eq:gapT}
\eeq
Here (and for all future terms we will consider), we imagine the scenario where the velocities and forward scattering interactions contained in $V$ in our Lagrangian are tuned such that this interaction is relevant \cite{fradkin1}; this can always be done\cite{Cano1}. As such we will consider the $\lambda = -\infty$ fixed point here, and for all future interactions. We can interpret Eq. \ref{eq:gapT} as a tunneling term or a condensation term. From the tunneling point of view, we can see that only processes that remove a $e^{i\phi_{t,1}}$ particle and add $n$ $e^{i \phi_{t,2}}$ particles (or vice versa) commute with Eq. \ref{eq:gapT}. From the condensation point of view, this term generates an expectation value for $\phi_{t,1} - n\phi_{t,2}$  when $\lambda \rightarrow -\infty$. From here on, we will refer to the specific interface given by Eq. \ref{eq:gapT}, and its associated null-vector, as the "traditional" interface, as it is the simplest and most well-studied.

\section{Stably Equivalent Interfaces}
While the traditional interface is the simplest, there are other natural interfaces to consider if we allow for edge "reconstruction" effects\cite{Wen9,Cano1}. We will now consider a situation where an unprotected, non-chiral pair of local particle channels, e.g., free electron channels on the edge of a fermionic Laughlin state, interact with our system. These states are not required by the topology, but can be generated by physical mechanisms on the edge of a system. One can think of this as either adding an additional quasi-1D layer with non-chiral local edge modes to the edge of our system, or perhaps weakening/removing gapping terms which have gapped out some pre-existing local edge modes in our system. Allowing for these channels puts us in a position to explore topological orders that are stably equivalent to one another\cite{Kitaev1,Cano1}, since they only differ by the addition of extra trivial sectors.

\emph{Confined interface--}
Let us add a single set of non-chiral channels to the $1/pn^2$ edge. For the $1/pn^2$ state this scenario can realized by using the K matrix
\beq
\bar{K}_{pn^2} = pn^2\oplus \Sigma
\label{eq:KRecon}
\eeq
where $\Sigma$ is a Pauli matrix. As a check for consistency, note that the ground state degeneracy on a torus $\det (\bar{K}_{pn^2}) = \det (K_{pn^2}) = pn^2$ is unchanged by the addition of $\Sigma$; the quasiparticle statistics are also unchanged. 
We will consider the cases where we use $\sigma^x$ (describing local bosons) for $p$ even, and $\sigma^z$ (describing local fermions) for $p$ odd. Using
\beq
W_{even} &=&  \begin{bmatrix}
    0  & 0 & -1  \\
    -1  & 0 & n \\
    -q  & 1 & -nq 
\end{bmatrix}\\ W_{odd} &=&  \begin{bmatrix}
    0  & 0 & 1  \\
    1+q  & -1 & nq \\
    q  & -1 & nq+n 
\end{bmatrix}
\label{eq:WMat}
\eeq
for $p = 2q$ even and $p = 2q+1$ odd respectively, we recast $\bar{K}_{pn^2}$ as  
\beq
K_{p,n^2} =  W^T \bar{K}_{pn^2} W = \begin{bmatrix}
    p  & -1 & 0  \\
    -1  & 0 & n \\
    0  & n & 0 
\end{bmatrix}.
\label{eq:kMatG}
\eeq
In Eq. \ref{eq:kMatG} it is manifest that the $1/pn^2$ state with local particles can be interpreted as a $1/p$ Laughlin state coupled to a $\mathbb{Z}_n$ gauge theory given by $n\sigma^x$, and where the $1/p$ state is sensitive to the discrete gauge flux. 

 Using this form, it is natural to consider a new interface which involves confining the emergent $\mathbb{Z}_n$ gauge field of the $1/pn^2$ Laughlin state. As we shall see, the quasiparticles that remain free after this will describe an effective Laughlin $1/p$ state. A completely gapped interface can then be created by coupling the anti-chiral $1/p$ state to the newly created effective chiral $1/p$ state. Because of the confinement of the $\mathbb{Z}_n$ gauge field, we will call this the "confined" interface. 

Formally, this is done as follows. The full interface is given by 
\beq
K_c = K_{-p}\oplus K_{p,n^2} = \begin{bmatrix}
	-p & 0 &0 & 0\\    
    0 & p  & -1 & 0  \\
   0 & -1  & 0 & n\\
   0 & 0  & n & 0 
\end{bmatrix},
\eeq 
with bosonic fields $\bm{\phi}_c = (\phi_{c,1},\phi_{c,2},\phi_{c,3},\phi_{c,4})$. The field $\phi_{c,1}$ corresponds to edge excitations of the anti-chiral $1/p$ state. As mentioned, $K_{p,n^2}$ describes an effective chiral $1/p$ state coupled to a $\mathbb{Z}_n$ gauge field, hence  $\phi_{c,2}$ corresponds to the edge excitations of the chiral $1/p$ state, $\phi_{c,3}$ corresponds to the charge excitations of the $\mathbb{Z}_n$ gauge field, and $\phi_{c,4}$ corresponds the conjugate fluxes of the $\mathbb{Z}_n$ gauge field. To confine the $\mathbb{Z}_n$ gauge field, the charge excitations can be condensed. This is done with the local term
\beq
H_C = \lambda \cos(n \phi_{c,3}),
\label{eq:gapC1}
\eeq
which corresponds to a null vector $\Lambda_{c,1} = (0,0,0,1)$.

After the charge condensation, the free excitations will be those that have unchanged self statistics after fusion with the condensed charged boson \cite{hung1}. A general excitation is given by $e^{i \bm{r} \cdot \bm{\phi}_{c}}$, where $\bm{r}$ is an integer valued vector. Let the charged boson be given by $e^{i \bm{b} \cdot \bm{\phi}_{c}}$ where $\bm{b} = (0,0,1,0)$. A free excitation will have $\bm{r}$ such that $(\bm{r} + \bm{b})K^{-1}_{c}(\bm{r} + \bm{b}) = \bm{r} K^{-1}_{pn^2}\bm{r}$. Solving for $\bm{r}$ gives that the free quasiparticles are generated by $e^{i \phi_{c,1}}$ and $e^{i \phi_{c,2}}$ (modulo fusion with $e^{i \bm{b} \cdot \bm{\phi}_{c}}$). These are the quasiparticle excitations of the anti-chiral $1/p$ edge and the effective chiral $1/p$ edge respectively. Explicit calculation of the self and mutual statistics of these quasiparticles confirms this. In terms of the original anyons of the $1/pn^2$ Laughlin state, the quasipartices of the effective chiral $1/p$ state are bound states composed of $n$ $1/pn^2$ anyons. 

The interface can be completely gapped by adding an additional back-scattering between the anti-chiral $1/p$ edge and the effective chiral $1/p$ state. This is done with the term
\beq
H_I = \lambda \cos(p\phi_{c,1} - p\phi_{c,2} + \phi_{c,3}),
\label{eq:gapC2}
\eeq
corresponding to the null vector $\Lambda_{c,2} = (1,1,0,0)$. The effect of Eq. \ref{eq:gapC1}, is to produce an expectation value for $\phi_{c,3}$, so Eq. \ref{eq:gapC2} is equivalent to the effective tunneling term
\beq
H_I = \lambda \cos(p\phi_{c,1} - p\phi_{c,2} + \langle \phi_{c,3} \rangle).
\label{eq:gapCSub}
\eeq
Since $\langle \phi_{c,3} \rangle$ only corresponds to a shift in the phase of in the tunneling term, it will be ignored (as will similar contributions) from now on. 
As a check for consistency, the null vectors $\Lambda_{c,1} $ and $\Lambda_{c,2}$ from Eqs. \ref{eq:gapC1} and \ref{eq:gapC2} satisfy the null vector criteria and primitivity conditions for $K_c = K_{-p}\oplus K_{p,n^2}$ \cite{Bar1}. Eqs. \ref{eq:gapC1} and \ref{eq:gapC2} thereby fully gap the interface and produce what we have called the confined interface.

\emph{Partially confined interface--} Following the same method as before, it is also possible to confine a $\mathbb{Z}_{n/m}$ gauge field if $n$ is divisible by $m$, i.e., $n/m \in \mathbb{Z}$. This can be thought of as confining a subgroup of the $\mathbb{Z}_n$ gauge field. This is done through the same procedure as confining the $\mathbb{Z}_n,$ gauge field but with the identification $p \rightarrow pm^2$, $n \rightarrow n/m$. Recasting the K-matrix in an analogous fashion to before we arrive at 
\beq
{K}_{pm^2,{n/m}^2} =  W^T K_{pn^2} W = \begin{bmatrix}
    pm^2  & -1 & 0  \\
    -1  & 0 & \frac{n}{m} \\
    0  & \frac{n}{m} & 0 
\end{bmatrix}.
\label{eq:kMatPG}
\eeq
The $1/pn^2$ state is thereby also equivalent to an effective $1/pm^2$ state coupled to a $\mathbb{Z}_{n/m}$ gauge field. An interface can be created by confining the $\mathbb{Z}_{n/m}$ gauge field, and then gapping out the anti-chiral $1/p$ state with the effective chiral $1/pm^2$ state with a traditional interface. 

This interface is described by the $K$ matrix
\beq
K_{pc} = K_{-p}\oplus K_{pm^2,n/m^2} = \begin{bmatrix}
	-p & 0 &0 & 0\\    
    0 & pm^2  & -1 & 0  \\
   0 & -1  & 0 & \frac{n}{m} \\
   0 & 0  & \frac{n}{m} & 0 
\end{bmatrix},
\eeq
with bosonic fields $\bm{\phi}_{pc} = (\phi_{pc,1},\phi_{pc,2},\phi_{pc,3},\phi_{pc,4})$. $\phi_{pc,1}$ corresponds to edge excitations of the anti-chiral $1/p$ state, $\phi_{pc,2}$ corresponds to the edge excitations of the effective chiral $1/pm^2$ state, $\phi_{pc,3}$ corresponds to the charge excitations of the $\mathbb{Z}_{n/m}$ gauge field, and $\phi_{pc,4}$ corresponds to the fluxes of the $\mathbb{Z}_{n/m}$ gauge field.

The $\mathbb{Z}_{n/m}$ gauge field is confined by the null vector $\Lambda_{pc,1} = (0,0,0,1)$, which generates the term
\beq
H_C = \lambda \cos(\frac{n}{m}\phi_{pc,3}).
\label{eq:gapP1}
\eeq
Using the same reasoning as for the confined interface, the free quasiparticles are generated by $e^{i \phi_{pc,1}}$ and $e^{i \phi_{pc,2}}$, which are the quasiparticles of the $1/p$ and effective $1/pm^2$ states respectively. In terms of the original anyons of the $1/pn^2$ state, the quasiparticles of the effective $1/pm^2$ state are given by bound states of $n/m$ $1/pn^2$ anyons. 

\begin{figure*}[t!]
\includegraphics[width=\linewidth]{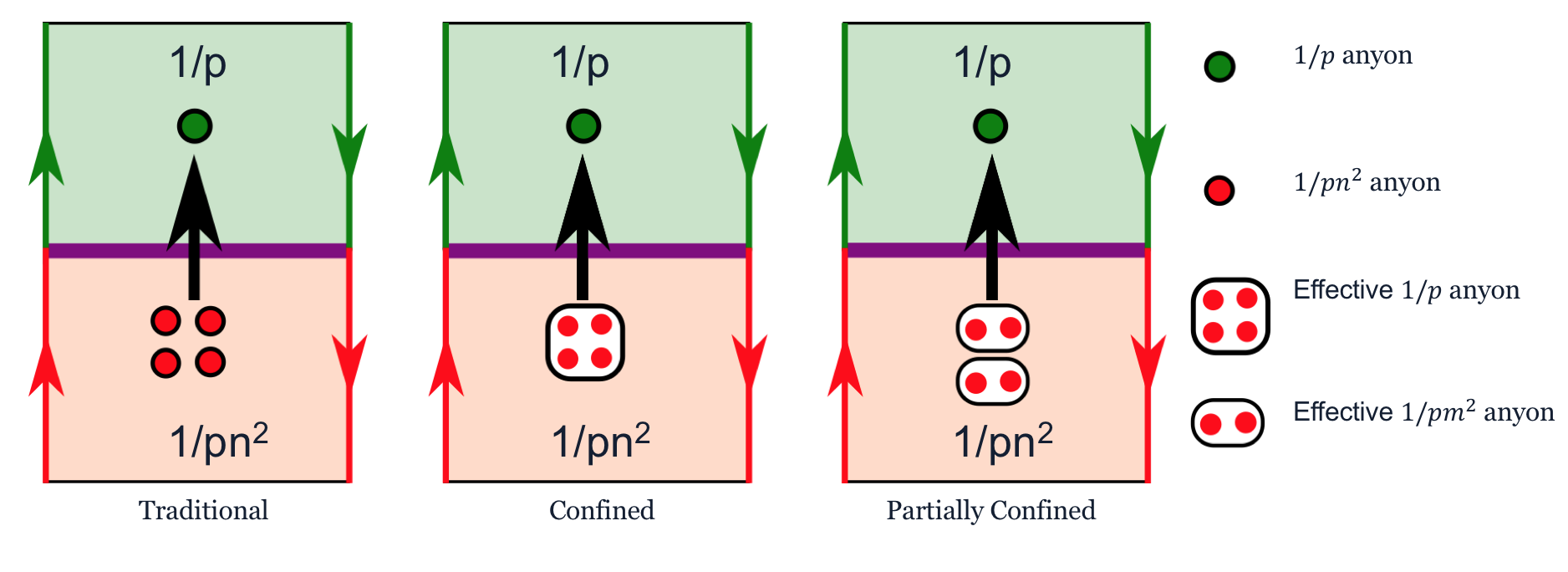}
\caption{Anyon tunneling between a chiral Laughlin $1/p$ (green) and an anti-chiral Laughlin $1/pn^2$ (red) state for different interfaces.
In the diagrams we have chosen $p=3$, $n = 4,$ and $m = 2$ as an explicit example.}
\label{fig:sDiagram}
\end{figure*}

The interface can be completely gapped by inducing back-scattering between the $1/p$ state and the effective $1/pm^2$ state. This is done with 
\beq
\nonumber H_I &=& \lambda \cos(pm\phi_{pc,1} - pm^2\phi_{pc,2} + \phi_{pc,3})\\
&\rightarrow& \lambda \cos(pm\phi_{pc,1} - pm^2\phi_{pc,2} + \langle \phi_{pc,3}\rangle),
\label{eq:gapP2}
\eeq
which corresponds to the null vector $\Lambda_{pc,2} = (m,1,0,0)$. $\Lambda_{pc,1}$ and $\Lambda_{pc,2}$ satisfy the null vector criteria and the primitivity condition, so the interface has been completely gapped. Because only part of the $\mathbb{Z}_n$ gauge field is confined, we shall refer to this interface as the "partially confined" interface. The partially confined interface can be seen as intermediate between the confined ($m = 1$), and traditional ($m = n$) interfaces.

Currently we have identified three types of interfaces which satisfy the null vector criteria and the primitivity condition. The traditional interface, the confined interface, and the partially confined interface. Besides their varied mathematical constructions we aim to find distinguishing physical characteristics. Below we will discuss the similarities and differences of these interfaces in regards to quasiparticle tunneling, ground state degeneracy, and parafermion zero modes where the interface intersects the vacuum.

\section{Tunneling Properties}
The tunneling properties for the three interfaces we are considering are summarized in Fig. \ref{fig:sDiagram}. As we shall show, these tunneling processes provide a physical way to differentiate these interfaces.

The traditional interface has previously been shown to act as an anyonic Andreev reflector\cite{santos2016}. This is because the tunneling process involves $n$ $e^{i\phi_{t,2}}$ quasiparticle and a single $e^{i\phi_{t,1}}$ quasiparticle (see Fig. \ref{fig:sDiagram} and Eq. \ref{eq:gapT}). As a result, moving a single $e^{i\phi_{t,2}}$ quasiparticle across the interface will result in a single $e^{i\phi_{t,1}}$ quasiparticle being transmitted to the $1/p$ side and $n-1$ $e^{-i \phi_{t,2}}$ (anti-)quasiparticles being reflected to the $1/pn^2$ side. Because of this the interface is a mod$(n)$ anyonic Andreev reflector. 

For the confined interface, the tunneling process involves a single $e^{i\phi_{c,2}}$ quasiparticle a single $e^{i\phi_{c,1}}$ quasiparticle (see Eq. \ref{eq:gapCSub}). The other possible quasiparticles are given by $e^{i \phi^2_{c,3}}$ and $e^{i \phi_{c,4}}$. Due to the condensate generated by Eq. \ref{eq:gapC1}, the $e^{i \phi_{c,3}}$ quasiparticles are absorbed into the condensate at the edge. Additionally, since $\phi_{c,3}$ and $\phi_{c,4}$ do not commute, the $e^{i \phi_{c,4}}$ quasiparticles are no longer well defined excitations at the interface, and thereby cannot tunnel to the other side. Hence, both the $e^{i \phi_{c,3}}$ and $e^{i \phi_{c,4}}$ quasiparticles are incapable of tunneling through the interface. Since the only tunneling process involves a single $e^{i\phi_{c,2}}$ quasiparticle a single $e^{i\phi_{c,1}}$ quasiparticle, it is clear that the confined interface is not a anyonic Andreev reflector. The absence of anyonic Andreev reflection provides a clear distinction between the confined interface and the traditional interface. 

For the partially confined interface, the tunneling process involves $m$ $e^{i\phi_{pc,2}}$ quasiparticle and a single $e^{i\phi_{pc,1}}$ quasiparticle (see Eq. \ref{eq:gapP2}). Moving a single $e^{i\phi_{pc,2}}$ quasiparticle across the interface thereby results in a single $e^{i\phi_{pc,1}}$ quasiparticle being transmitted to the $1/p$ side and $m-1$ $e^{i\phi_{pc,2}}$ (anti-)quasiparticles being reflected to the $1/pn^2$ side. For the same reasons as in the confined interface, all other excitations cannot tunnel into the $1/p$ edge. So the partially confined interface is a mod$(m)$ anyonic Andreev reflector. We can conclude that the possible tunneling mechanisms across the interface can help distinguish the interface types for a given fixed set of topologically ordered states on either side of the interface.

\section{Ground State Degeneracy}

In order to characterize the interfaces using ground state degeneracy (GSD) we will take our system and fold it at the interface to create a bi-layer system where the interface is now a single (bi-layer) edge. The interface terms will then gap this edge. This bi-layer can then be put on a manifold with boundary and the GSD can be calculated for the various types of gapped boundaries. 

First we consider the case where the topological order ($K$-matrix) is put on a disk, and the boundary of the disk is gapped by one of the previously determined interfaces. For a general $2N\times 2N$ $K$-matrix on a disk, which has its edge gapped by a backscattering term of the form of Eq. \ref{eq:gapGe}, the GSD is given by gcd$(N\times N$ minors of  $M_d)$, where $M_d = [\Lambda_1 ... \Lambda_N]$ \cite{Levin3}. Due to the primitivity condition, we have applied for our null vectors, the GSD on a disk must be $1$ for all of the interfaces we constructed. Explicitly, for the traditional interface $M_d = (n,1)^T$, and gcd$(1\times 1$ minors of  $M_d) = 1$. So the ground state on a disk is indeed unique. Thus, a disk geometry does not help distinguish the interfaces.

Now, we will now consider the more interesting case where we put the bilayer on a cylinder and use a given gapping term on both edges. In this case the GSD is gcd$(2N\times 2N$ minors of  $M_c)$, where \cite{Levin3}
\beq
M_c = \begin{bmatrix}
    K\Lambda_1  & ...  &  K\Lambda_N & 0 & ... & 0 \\
    \Lambda_1  & ...  &  \Lambda_N  & \Lambda_1  & ...  &  \Lambda_N 
\end{bmatrix}.
\eeq
For the traditional interface on both edges of the cylinder there are $pn$ ground states. Using either the confined or partially confined interfaces on both edges the GSD is also $pn$. At first sight we find that the GSD is identical for all three interfaces, but we can look more closely to find some distinguishing characteristics.

Let us consider a helpful perspective for the GSD. The ground state degeneracy can be understood by examining the edge physics on the cylinder. Consider a cylinder where both edges are gapped using a traditional interface. At each end $\phi_{t,1}-n\phi_{t,2}$ will have an expectation value corresponding to a minimum of Eq. \ref{eq:gapT}. There is a conserved anyon charge density given by $\rho_t = \partial_x (\phi_{t,1} - n\phi_{t,2})/2\pi$. The total anyon charge on the cylinder is then
\beq
Q_{t} = (\langle \phi_{t,1}-n\phi_{t,2} \rangle_L + \langle \phi_{t,1}-n\phi_{t,2} \rangle_R)/2\pi
\eeq
where $\langle ... \rangle_{L/R}$ is the expectation value on the left and right edges of the cylinder. In a physical system there should be no net anyon charge, i.e., no net fractionalized particles in the system. As a result the ground state must satisfy $\langle \phi_{t,1}-n\phi_{t,2} \rangle_L = -\langle \phi_{t,1}-n\phi_{t,2} \rangle_R$. Since $\langle \phi_{t,1}-n\phi_{t,2} \rangle_{L/R}$ are bound to the minima of Eq. \ref{eq:gapT}, there are $pn$ unique possible ways to satisfy this constraint. These are the $pn$ ground states we calculated earlier. The ground states are thereby identified with the different minima of the term Eq. \ref{eq:gapT} which gaps the edges. This agrees with earlier work on ground state degeneracy in systems with gapped boundaries\cite{Wen10}.

\begin{figure}[h]
\includegraphics[width=\linewidth*3/4]{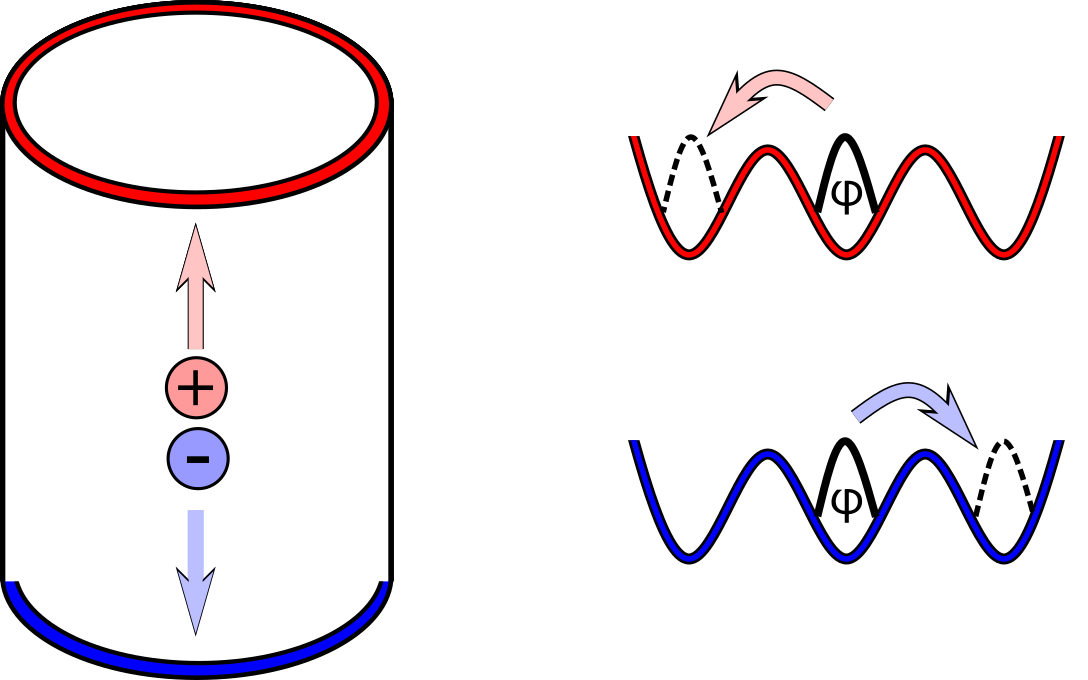}
\caption{The process of changing ground states by adding and removing charge from opposite ends of a cylinder with gapped edges. This changes the expectation value of the fields pinned at the edges. }
\label{fig:GSFig}
\end{figure}

Using this identification, we can consider the process of moving between different ground states. Physically, this involves adding a quantized amount of anyon charge to the left edge and removing the same amount anyon charge from right edge. This process will move $\langle \phi_{t,1}-n\phi_{t,2} \rangle_L = -\langle \phi_{t,1}-n\phi_{t,2} \rangle_R$ to a different minima of Eq. \ref{eq:gapT}. So changing ground states corresponds to moving between different minima of Eq. \ref{eq:gapT} as shown in Fig \ref{fig:GSFig}. The action of moving between different minima gives the $pn$ ground states the group structure of the cyclic group $\mathbb{Z}_{pn}$.

The group structure of the ground states for a cylinder with either confined or partially confined interfaces on both edges can be found with this perscription. For the \emph{confined} interface at both ends, the fields $\phi_{c,3}$ and $\phi_{c,1}-\phi_{c,2}$ will both have expectation values due to Eq. \ref{eq:gapC1} and Eq. \ref{eq:gapCSub} respectively. There are then two conserved anyon charges given by
\beq
\nonumber Q_{c,1} &=& (\langle\phi_{c,3}\rangle_L + \langle \phi_{c,3}\rangle_R)/2\pi\\
Q_{c,2} &=& (\langle \phi_{c,1}-\phi_{c,2} \rangle_L + \langle \phi_{c,1}-\phi_{c,2} \rangle_R)/2\pi.
\eeq
Both of these charges should be zero for a physical system, so $Q_{c,1} = Q_{c,2} = 0$. Since $\langle\phi_{c,3}\rangle$ is pinned to a minimum of Eq. \ref{eq:gapC1} there are $n$ ways to satisfy $Q_{c,1} = 0$. Similarly, there are $p$ ways to satisfy  $Q_{c,2} = 0$ due to  Eq. \ref{eq:gapCSub}. In total there are $pn$ ground states, in agreement with the earlier calculation.

As in the traditional interface, adding and removing anyon charge changes the ground state. However, for the confined interface there are two distinct types of anyon charge $Q_{c,1}$ and $Q_{c,2}$. Adding and removing $Q_{c,1}$ corresponds to moving the value of $\langle\phi_{c,3}\rangle_L = - \langle \phi_{c,3}\rangle_R$ to a different minima of Eq. \ref{eq:gapC1}, while adding and removing $Q_{c,2}$ corresponds to moving the value of $\langle \phi_{c,1}-\phi_{c,2} \rangle_L  = - \langle \phi_{t,1}-\phi_{c,2} \rangle_R$ to a different minima of Eq. \ref{eq:gapCSub}.
Since moving between the $n$ minima of Eq. \ref{eq:gapC1} and the $p$ minima of Eq. \ref{eq:gapCSub} are independent actions, the group structure of the ground states is $\mathbb{Z}_p \times \mathbb{Z}_{n}$. Although the number of ground states is the same for both the traditional and confined interfaces, we find that the group structure of the ground states is in fact different. 

For the \emph{partially confined} interface at both ends, the fields $\phi_{pc,3}$ and $\phi_{pc,1}-m\phi_{pc,2}$ will both have expectation values due to Eq. \ref{eq:gapP1} and Eq. \ref{eq:gapP2} respectively. There are again two conserved anyon charges given by
\beq
 Q_{pc,1} &=& (\langle\phi_{pc,3}\rangle_L + \langle \phi_{pc,3}\rangle_R)/2\pi\\\nonumber
Q_{pc,2} &=& (\langle \phi_{pc,1}-m\phi_{pc,2} \rangle_L + \langle \phi_{pc,1}-m\phi_{pc,2} \rangle_R)/2\pi.
\eeq
 Due to Eqs. \ref{eq:gapP1} and \ref{eq:gapP2} there are $n/m$ ways to satisfy $Q_{pc,1} = 0$, and $pm$ ways to satisfy $Q_{pc,2} = 0$ leading to $pn$ ground states. Following the same logic as was used for the confined interface, the group structure of the ground states for the partially confined interface is $\mathbb{Z}_{pm} \times \mathbb{Z}_{n/m}$.
 
\section{Parafermion Zero Modes}

In earlier work it was shown that for the traditional interface between the $1/p$ and $1/pn^2$ states,  there are parafermion zero modes if the interface terminates at the vacuum.\cite{cano2014,santos2016,santos2018} We want to understand the nature of the possible parafermion modes for the confined and partially confined interfaces. 

\begin{figure}[h]
\includegraphics[width=\linewidth*3/4]{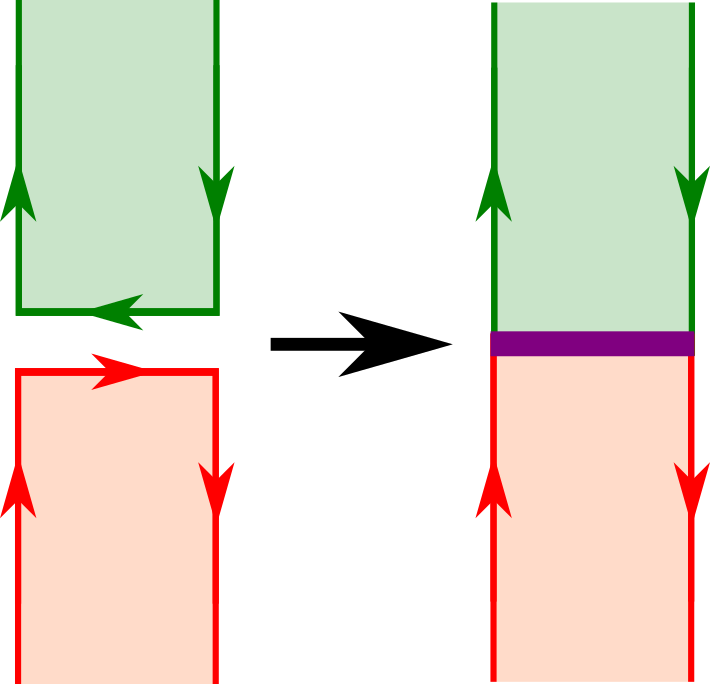}
\caption{Combining two semi-infinite topologically ordered strip into a single infinite strip by using a gapped interface that terminates at the vacuum (purple).}
\label{fig:iDiagram3}
\end{figure}

\begin{figure}[h]
\includegraphics[width=\linewidth*4/5]{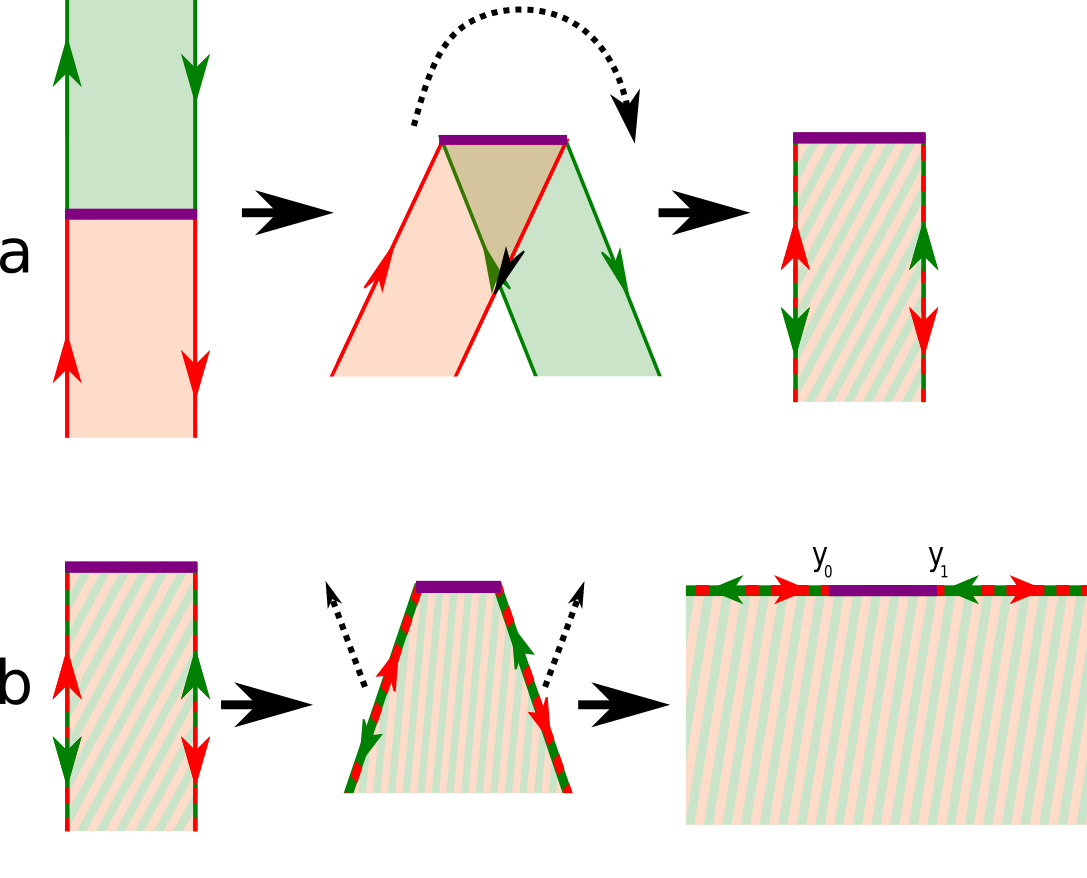}
\caption{(a)Folding an infinite topologically ordered semi-infinite strip into a semi-infinite strip. (b) Deforming the semi-infinite strip into a half-plane. }
\label{fig:iDiagram2}
\end{figure}

To begin, we reconsider the traditional interface and re-derive the existence of the parafermion zero modes. Additionally, we will present the parafermion zero modes in terms of an "order" and "disorder" operator. This is done by considering two semi-infinite strips, one with $1/p$ topological order, and one with $1/pn^2$ topological order, which are combined into a single infinite strip by using a traditional interface (as shown in Fig \ref{fig:iDiagram3}). The interface between the two topological orders thereby terminates at the vacuum while the two edges of the strip remain gapless. In order to consider only a single edge, we will fold the infinite strip along the interface leading to a semi-infinite strip with a single edge and topological order given by the $K$ matrix $K_t$ (see Fig \ref{fig:iDiagram2}a). The section of the edge where the strip was folded will be gapped by the interface term used to join the $1/p$ and $1/pn^2$ states (purple in Fig \ref{fig:iDiagram2}a). The rest of the edge will remain gapless (red and green stripes in Fig \ref{fig:iDiagram2}a). 

Topologically, a semi-infinite strip is the same as a half-plane, so it will be useful to deform the strip into half plane as is done in Fig \ref{fig:iDiagram2}b. The edge of the half-plane then consists of a finite gapped section (due to the interface term) and gapless non-chiral modes to the left and right. Defining the gapped section of the edge as running between $y_{0}$ and $y_{1}$ ($y_{0}<y_{1}$), the interaction along the edge is given by  
\beq
H_I = \lambda\cos(pn\phi_p - pn^2 \phi_{pn^2}) \Theta(y-y_0)\Theta(y_1-y),
\label{eq:gapT2}
\eeq
where $\Theta$ is a step function. 

At the ends of the gapped section of the edge ($y_0$ and $y_1$), we define the operators 
\beq
\nonumber \chi_{t,0} &=& e^{i\bm{l}_o \cdot \bm{\phi}_t(y_0+\epsilon)}e^{i\bm{l}_d\cdot\bm{\phi}_t(y_0 - \epsilon)}\\
\nonumber \chi_{t,1} &=& e^{i\bm{l}_d\cdot\bm{\phi}_t(y_1 + \epsilon)}e^{i\bm{l}_o \cdot \bm{\phi}_t(y_1-\epsilon)}\\
\nonumber \bm{l}_o &=& (p,-pn)\\ 
\bm{l}_d &=& (1/2n,1/2),
\eeq
where $\bm{\phi}_t$ are the previously defined bosons of the $K_t$ edge and $\epsilon = 0^+$ (the subscripts $o$ and $d$ will be connected to the order/disorder operator language below). Both $\chi_{t,0}$ and $\chi_{t,1}$ commute with the Hamiltonian, and $\chi_{t,1} \chi_{t,0} = \chi_{t,0} \chi_{t,1} e^{i\frac{2\pi}{n}}$. Generalizing this to multiple interfaces and operators $\chi_{t,i}$, the commutation relationship are given by $\chi_{t,i} \chi_{t,j} = \chi_{t,j} \chi_{t,i} e^{-i\frac{2\pi}{n}sgn(i-j)}$ which is the parafermion algebra. 

This construction of the $\chi_t$ operators has a natural interpretation in terms of order and disorder operators. For the traditional interface on the half plane, the parafermion operator $\chi_{t,0}$ is composed of $e^{i\bm{l}_o \cdot \bm{\phi}_t(y_0+\epsilon)}$, and $e^{i\bm{l}_d\cdot\bm{\phi}_t(y_0 - \epsilon)}$. Due to Eq. \ref{eq:gapT2} $e^{i\bm{l}_o \cdot \bm{\phi}_t(y_0+\epsilon)}$ will have an expectation value. As such we identify $e^{i\bm{l}_o \cdot \bm{\phi}_t(y_0+\epsilon)}$ as an order operator. Since $[\bm{l}_o\cdot\bm{\phi}_t(y_i),\bm{l}_d\cdot\bm{\phi}_t(y_j)] = -i\frac{\pi}{n}{\rm{sgn}}(y_i - y_j)$, we identify $e^{i\bm{l}_d\cdot\bm{\phi}_t}$ as a $\mathbb{Z}_n$ disorder which creates kinks in the value of $e^{i\bm{l}_o \cdot \bm{\phi}_t}$. Both the order and disorder operators self commute, and commute with the Hamiltonian. Because of this, $\chi_{t,0}$ is necessarily a zero energy parafermion operator. The same argument holds for $\chi_{t,1}$.

This construction also shows that the parafermion zero mode at the edge of the interface has quantum dimension $\sqrt{n}$. Explicitly, the operators 
\beq
\nonumber O^{-}_0 &=& e^{i\bm{l}_o \cdot \bm{\phi}_t(y_0+\epsilon)}\\
\nonumber O^{-}_1 &=& e^{i\bm{l}_o \cdot \bm{\phi}_t(y_1-\epsilon)}\\ 
O^{+}_0 &=& e^{i\bm{l}_d\cdot\bm{\phi}_t(y_0 - \epsilon)} e^{i\bm{l}_d\cdot\bm{\phi}_t(y_1 + \epsilon)}
\eeq
can also be constructed from the order operators $e^{i\bm{l}_o \cdot \bm{\phi}_t}$ and disorder operators $e^{i\bm{l}_d\cdot\bm{\phi}_t}$ used to make $\chi_t$. Because the order and disorder operators individually commute with the Hamiltonian, the $O^{\pm}$ operators also commute with the Hamiltonian.
Assuming $\bm{\phi}_t$ vanishes at $\pm \infty$, the $O^{\pm}$ operators can be rewritten as  
\beq
O^{-}_{0/1} &=& \exp[i \int_{R^-_{0/1}} dy' \partial_{y'}(\bm{l}_o \cdot \bm{\phi}_t)]\\\nonumber
O^{+}_0 &=& \exp[i \int_{R^+_0} dy' \partial_{y'} (\bm{l}_d\cdot\bm{\phi}_t)],
\eeq
where $R^-_0 = (-\infty, y_{0}+\epsilon)$, $R^+_0 = (y_{0}-\epsilon,y_{1}+\epsilon)$, and $R^-_1 = (y_{1}-\epsilon,\infty)$. These operators satisfy the algebra $O^{-}_0 O^{+}_0 = O^{+}_0 O^{-}_0 e^{-i\frac{2\pi}{n}}$, $O^{-}_1 O^{+}_0 = O^{+}_0 O^{-}_1 e^{i\frac{2\pi}{n}}$, and $O^{-}_0 O^{-}_1 = O^{-}_1 O^{-}_0$. 

Since these operators commute with the Hamiltonian, we can label our ground states in the  $O^{-}_{0/1}$ basis, and use $O^{+}_0$ as a ladder operator. This algebra indicates that for the traditional interface, the ground states must come in multiples of $n$. This can be generalized to multiple interfaces which would involve multiple parafermion operators. The analogous operators will then be $O^{-}_i$, and $O^{+}_j$ with $O^{-}_i O^{+}_j = O^{+}_j O^{-}_i e^{i\frac{2\pi}{n}(\delta_{j,i-1} - \delta_{j,i})}$. So the ground states must come in mutliples of $n^k$ for $k$ interfaces on the half plane \cite{Hughes1}. For a disk geometry, the ground states will come in multiples of $n^{k-1}$ for $k$ interfaces since the operators $O^{\pm}_i$ are no longer independent. In both cases, each additional interface (beyond possibly the first) adds $n$ ground states. In summary, the construction of the zero energy parafermion operator $\chi$ from the order and disorder operators is equivalent to the existence of the operators $O^{\pm}$ which commute with the Hamiltonian, and generate $n$ ground states per interface. A pair of parafermion zero modes at the ends of an interface thereby indicate $n$ ground states, and so each parafermion zero mode has a quantum dimension of $\sqrt{n}$ as claimed. 

\begin{table*}
\caption{Interface Results} 
\centering 
\begin{tabular}[t]{|m{18mm}|m{38mm}|m{25mm}|m{35mm}|m{45mm}|} 
\hline\hline 
Name &  $K$ matrix and \newline Null Vectors & Tunneling \newline Properties & Ground State \newline Group Structure on a \newline Cylinder  & Edge Parafermions \newline $e^{i\bm{l}_o\cdot\bm{\phi}}e^{i \bm{l}_d\cdot\bm{\phi}}$ \\  
\hline \hline 
Traditional & $K_t = \begin{bmatrix}
    -p  & 0   \\
    0  & pn^2
\end{bmatrix}$ \newline $\Lambda_t = (n,1)$&mod($n$) anyon\newline Andreev reflection &$\mathbb{Z}_{pn}$& $\mathbb{Z}_n$ \newline $\bm{l}_o = (p,-pn)$\newline $\bm{l}_d = (\frac{1}{2n},\frac{1}{2})$\\\hline 
Confined & $K_c = \begin{bmatrix}
	-p & 0 &0 & 0\\    
    0 & p  & -1 & 0  \\
   0 & -1  & 0 & n\\
   0 & 0  & n & 0 
\end{bmatrix}$ \newline $\Lambda_{c,1} = (0,0,0,1)$  \newline $\Lambda_{c,2} = (1,1,0,0)$ & no anyon \newline Andreev reflection &  $\mathbb{Z}_n \times \mathbb{Z}_p$   & $\mathbb{Z}_n$ \newline $\bm{l}_o = (0,0,1,0)$\newline $\bm{l}_d = (0,\frac{1}{n},0,-1)$\\\hline 
Partially \newline Confined & $K_{pc} = \begin{bmatrix}
	-p & 0 &0 & 0\\    
    0 & pm^2  & -1 & 0  \\
   0 & -1  & 0 & \frac{n}{m} \\
   0 & 0  & \frac{n}{m} & 0 
\end{bmatrix}$ \newline $\Lambda_{pc,1} = (0,0,0,1)$  \newline $\Lambda_{pc,2} = (m,1,0,0)$& mod($m$) anyon \newline Andreev reflection &  $\mathbb{Z}_{n/m} \times \mathbb{Z}_{pm}$ & $\mathbb{Z}_{n/m}$ \newline $\bm{l}_o = (0,0,1,0)$\newline $\bm{l}_d = (0,\frac{m}{n},0,-1)$\newline $\mathbb{Z}_{m}$ \newline $\bm{l}_o = (p,-pm,0,0)$\newline $\bm{l}_d = (\frac{1}{2m},\frac{1}{2},0,0)$ \\\hline 

\hline 
\end{tabular}
\label{table:resultsTable} 
\caption*{The summary of results for the 3 different gapped interfaces. }
\end{table*}

The order and disorder operator picture of the parafermion operators gives a general procedure for constructing $\mathbb{Z}_n$ parafermion zero modes. For a given interface, we can identify an order operator, along with a corresponding $\mathbb{Z}_n$ disorder operator which creates kinks in the order operator. Formally this is a pair of operators $e^{i\phi_o}$ and $e^{i\phi_d}$ satisfying $[\phi_o(y),\phi_d(y')] = -i\frac{\pi}{n}{\rm{sgn}}(y - y')$. Provided that both the order and disorder operators are bosonic, the product of these terms at a given interface will necessarily be a $\mathbb{Z}_n$ parafermion. As we have shown, if the order and disorder operators also commute with the Hamiltonian at the edges of the gapped interface, there are $\mathbb{Z}_n$ parafermion zero modes with quantum dimension $\sqrt{n}$. Parafermions for the confined and partially confined interfaces can now be constructed in terms of order and disorder operators. For simplicity, we will continue to use the half plane geometry of Fig. \ref{fig:iDiagram2} with a gapped edge between $y_0$ and $y_1$, and gapless modes everywhere else. 

For the confined interface, the region between $y_0$ and $y_1$ of the half plane are gapped with Eqs. \ref{eq:gapC1} and \ref{eq:gapCSub}. The parafermion operators are
\beq
\nonumber \chi_{c,0} &=& e^{i\bm{l}_o \cdot \bm{\phi}_c(y_0+\epsilon)}e^{i\bm{l}_d\cdot\bm{\phi}_c(y_0 - \epsilon)}\\
\nonumber\chi_{c,1} &=& e^{i\bm{l}_d\cdot\bm{\phi}_c(y_1 + \epsilon)}e^{i\bm{l}_o \cdot \bm{\phi}_c(y_1-\epsilon)}\\
\nonumber\bm{l}_{o} &=& (0,0,1,0)\\
\bm{l}_{d} &=& (0,1/n,0,-1),
\eeq
where $\bm{\phi}_c$ are the bosons of the $K_c$ edge. $\chi_{c,0}$ and $\chi_{c,1}$ commute with the Hamiltonian and $\chi_{c,1} \chi_{c,0} = \chi_{c,0} \chi_{c,1} e^{i\frac{2\pi}{n}}$, so they are $\mathbb{Z}_n$ parafermion zero modes with quantum dimension $\sqrt{n}$. It is worth noting that all fields in the parafermion operator originate from the $1/pn^2$ topological order. As a result, the $\mathbb{Z}_n$ parafermions will remain even if tunneling with the $1/p$ state (Eq. \ref{eq:gapCSub}) is removed.

For the partially confined interface, the region of the half plane between $y_0$ and $y_1$ is gapped by Eqs. \ref{eq:gapP1} and \ref{eq:gapP2}. As we shall show there are two independent sets of parafermion zero modes in this case. The first set of parafermions is given by
\beq
\nonumber \chi_{pc,0} &=& e^{i\bm{l}_o \cdot \bm{\phi}_{pc}(y_0+\epsilon)}e^{i\bm{l}_d\cdot\bm{\phi}_{pc}(y_0 - \epsilon)}\\
\nonumber\chi_{pc,1} &=& e^{i\bm{l}_d\cdot\bm{\phi}_{pc}(y_1 + \epsilon)}e^{i\bm{l}_o \cdot \bm{\phi}_{pc}(y_1-\epsilon)}\\
\nonumber\bm{l}_{o} &=& (0,0,1,0)\\
\bm{l}_{d} &=& (0,m/n,0,-1),
\eeq
where $\bm{\phi}_{pc}$ are the bosons of the $K_{pc}$ edge. $\chi_{pc,0}$ and $\chi_{pc,1}$ commute with the Hamiltonian and $\chi_{pc,1} \chi_{pc,0} = \chi_{pc,0} \chi_{pc,1} e^{i\frac{2\pi m}{n}}$, and so they are $\mathbb{Z}_{n/m}$ parafermion zero modes with quantum dimensions $\sqrt{n/m}$. The second set of parafermions is given by 
\beq
\nonumber \gamma_{pc,0} &=& e^{i\bm{l}_o \cdot \bm{\phi}_{pc}(y_0+\epsilon)}e^{i\bm{l}_d\cdot\bm{\phi}_{pc}(y_0 - \epsilon)}\\
\nonumber\gamma_{pc,1} &=& e^{i\bm{l}_d\cdot\bm{\phi}_{pc}(y_1 + \epsilon)}e^{i\bm{l}_o \cdot \bm{\phi}_{pc}(y_1-\epsilon)}\\
\nonumber\bm{l}_o &=& (p,-mp,0,0)\\
\bm{l}_d &=& (1/2m,1/2,0,0).
\eeq
Both $\gamma_{pc,0}$ and $\gamma_{pc,1}$ commute with the Hamiltonian and $\gamma_{pc,1} \gamma_{pc,0} = \gamma_{pc,0}\gamma_{pc,1}e^{i\frac{2\pi}{m}}$, and so they are $\mathbb{Z}_m$ parafermions zero modes with quantum dimensions $\sqrt{m}$. So there are independently both $\mathbb{Z}_m$ and $\mathbb{Z}_{n/m}$ parafermion zero modes for the partially confined interface. Importantly, the combination of $\mathbb{Z}_{m}$ and $\mathbb{Z}_{n/m}$ parafermions is not necessarily equivalent to $\mathbb{Z}_{n}$ parafermions, since $\mathbb{Z}_{m} \times \mathbb{Z}_{n/m}$ does not necessarily equal $\mathbb{Z}_n$.

\section{Conclusion} 
We conclude the analysis of the three interfaces with a summary of the results shown in Table \ref{table:resultsTable}. Despite the simple structure of the $1/p$ and $1/pn^2$ edges there are actually a variety of ways to create interfaces between them. This can lead to concrete differences in the tunneling, ground state, tunneling, and parafermion zero mode structures. These calculations may be of experimental interest due to the ability to realize and tune quantum hall edge interfaces in, e.g., bilayer graphene systems\cite{Fallahazad1}. In particular, this may allow for observation of parafermion zero modes which can be used for topological quantum computing\cite{nayak2008}. 

Compared to the traditional interface, the new interfaces discussed here require additional local particles. These local particles may naturally accompany certain edges due to reconstruction effects. These new types of interfaces may also provide insight into interfaces between Abelian and non-Abelian topological orders, which are related by gauging anyonic symmetries. For example, we can consider the Abelian deconfined $\mathbb{Z}_2$ gauge theory (toric code) and the non-Abelian double Ising topological orders, which are related by gauging the $e\leftrightarrow m$ anyonic symmetry of the $\mathbb{Z}_2$ gauge theory\cite{barkeshli2014,Teo1}. It may then be possible to create a generalized confined interface by condensing the $e\leftrightarrow m$ charged particle on the double Ising edge, and then gapping out the two copies of $\mathbb{Z}_2$ gauge theories. An interface for this system has been previously determined in Ref. \onlinecite{Wen6}, and we believe that this interface seems to correspond to a generalized confined interface. We leave these questions to future work.

\section*{Acknowledgements}
We thank L. Santos and M. Lapa for useful conversations. TLH acknowledges support from the US National Science Foundation under grant DMR 1351895-CAR.

%

\bibliographystyle{apsrev4-1}

\end{document}